\documentclass{article} 
\usepackage{iclr2022_conference,times}

\usepackage{amsmath,amsfonts,bm}

\def\eqref#1{equation~\ref{#1}}

\def\1{\bm{1}}

\DeclareMathAlphabet{\mathsfit}{\encodingdefault}{\sfdefault}{m}{sl}
\SetMathAlphabet{\mathsfit}{bold}{\encodingdefault}{\sfdefault}{bx}{n}

\usepackage{hyperref}
\usepackage{url}

\usepackage[utf8]{inputenc} 
\usepackage[T1]{fontenc}    
\usepackage{hyperref}       
\usepackage{url}            
\usepackage{booktabs}       
\usepackage{amsfonts}       
\usepackage{nicefrac}       
\usepackage{microtype}      
\usepackage{braket}
\usepackage{float}
\usepackage{amsmath}
\usepackage{amsthm}
\usepackage{graphicx}
\usepackage{hyperref}
\usepackage{mathtools}
\usepackage[font=small,skip=8pt]{caption}
\usepackage[export]{adjustbox}
\usepackage{duckuments}
\usepackage{xparse}
\usepackage{xcolor}         
\usepackage{soul}
\makeatletter
\NewDocumentCommand{\sotwo}{O{blue}O{blue}+m}
    {%
        \begingroup
        \setulcolor{#1}%
        \setul{-.5ex}{.4pt}%
        \def\SOUL@uleverysyllable{%
            \rlap{%
                \color{#2}\the\SOUL@syllable
                \SOUL@setkern\SOUL@charkern}%
            \SOUL@ulunderline{%
                \phantom{\the\SOUL@syllable}}%
        }%
        \ul{#3}%
        \endgroup
    }
\makeatother

\newtheorem{lemma}{Lemma}

\title{Image Compression and Classification Using Qubits and Quantum Deep Learning}

\author{
  Ali Mohsen\thanks{alternatively \texttt{ahm302@nyu.edu}} \\
  Department of Computer Science\\
  Stanford University\\
  Stanford, CA 94305 \\
  \texttt{ahm302@stanford.edu}\\
  \And
  \textbf{Mo Tiwari }\\
  Department of Computer Science\\
  Stanford University\\
  Stanford, CA 94305 \\
  \texttt{motiwari@stanford.edu} \\
}

\begin{document}

\maketitle

\begin{abstract}
\label{section:abstract}
Recent work suggests that quantum machine learning techniques can be used for classical image classification by encoding the images in quantum states and using a quantum neural network for inference.
However, such work has been restricted to very small input images, at most $4 \times 4$, that are unrealistic and cannot even be accurately labeled by humans.
The primary difficulties in using larger input images is that hitherto-proposed encoding schemes necessitate more qubits than are physically realizable. 
We propose a framework to classify larger, realistic images using quantum systems. 
Our approach relies on a novel encoding mechanism that embeds images in quantum states while necessitating fewer qubits than prior work.
Our framework is able to classify images that are larger than previously possible, up to $16 \times 16$ for the MNIST dataset on a personal laptop, and obtains accuracy comparable to classical neural networks with the same number of learnable parameters. 
We also propose a technique for further reducing the number of qubits needed to represent images that may result in an easier physical implementation at the expense of final performance.
Our work enables quantum machine learning and classification on classical datasets of dimensions that were previously intractable by physically realizable quantum computers or classical simulation.\footnote{The repository with the code required to run and reproduce the experiments in this paper can be found in the ~\nameref{section:reproducibility} section}

\end{abstract}

\section{Introduction}
\label{section:introduction}
In the past decade, deep learning has been remarkably successful on a wide variety of classical learning tasks (\cite{DeepLearning, LeCun}). In parallel, quantum computing (QC) has long promised dramatic increases in computational power over classical computers, culminating in a recent demonstration of quantum supremacy in a machine with $53$ programmable qubits in \cite{Sycamore1}. However, even these quantum systems are already approaching the limits of classical simulability by the world's largest traditional supercomputers (\cite{Sycamore2, Sycamore3}). The power of quantum computation suggests that quantum analogues of deep learning models like feedforward neural network may outperform their classical counterparts, especially when the data is inherently quantum (\cite{QNN2, QNN3, Farhi}). 

In this paper, we use a quantum neural network (QNN) to classify the MNIST dataset of handwritten digits (\cite{MNIST}). Prior work has been restricted to only highly-compressed, rather unrealistic input images due to their inefficient encoding schemes that are injective maps from classical images to their corresponding pure quantum states. These frameworks have used input images of a maximum resolution of $4 \times 4$, which is too coarse even for humans to provide accurate labels (see Figure \ref{fig:digit}). For larger images, injectivity would necessitate the same number of qubits as bits that are present in the original image. However, recent advances in quantum machine learning (QML) on classical data, such as the Flexible Representation of Quantum Images (FRQI) in \cite{FRQI}, have demonstrated that quantum wavefunctions can utilize quantum entanglement to encode \textit{classical} data using exponentially less qubits than their corresponding classical representation in bits. In this paper, we use the FRQI method to embed the input images in fewer-qubit systems. This approach necessitates a novel QNN architecture for classification, which we describe in Section \ref{section:methods}.   The input to our model are images of resolution up to $16\times16$ whose quantum encoding only requires $8$ qubits ($6$ for pixel locations, $1$ for color, and $1$ for readout). To the best of our knowledge, our work is the first to propose a data encoding scheme and QNN that can be used to classify \textit{realistic} images.


\textbf{Our main contributions are as follows:}
\begin{enumerate}
    \item We provide a novel construction to compress images and encode them in their FRQI states. Our construction uses only 2-qubit gates, which permits its use in common quantum machine learning packages such as Cirq and Tensorflow Quantum (TFQ) and may be of independent interest (\cite{Cirq,TFQ}).
    \item We propose a new QNN layers, CRADL and CRAML, which we use in a model trained with the images' FRQI states as input. 
    \item We show that our trained QNN achieves accuracy comparable to classical models with the same number of parameters.
    \item We propose a novel technique to further compress black and white images, and study the scaling behavior of our model with the extent of image compression.
\end{enumerate}

\textbf{Organization: }In Section \ref{section:preliminaries}, we provide a brief review of the formalism of quantum computation. In Section \ref{section:relatedwork}, we provide an overview of related work and motivate our study. In Section \ref{section:formalsetting}, we describe our dataset and how each image is encoded in a quantum state. In Section \ref{section:methods}, we describe the prescription for using a quantum neural network to obtain a classification prediction for a given image and describe the model we use for classification. In Section \ref{section:results}, we present our results. We conclude with Section \ref{section:conclusion} and a description of future work in Appendix \ref{section:futurework}.

\section{Preliminaries}
\label{section:preliminaries}

\subsection{Quantum Computing}
\label{subsection:quantumcomputing}

Here we provide a brief review of quantum computation. For a more detailed reference and an interactive coding tutorial, we refer the reader to \cite{NielsenChuang} and \cite{qiskit} respectively.


In quantum computation, the basic unit of information is a two-state quantum mechanical (QM) system called a qubit; the two states are traditionally written $\ket{0}$ and $\ket{1}$. A qubit can be in either of these two states, as well as a quantum superposition of these states, formally written as a wavefunction $\ket{\psi}=a_0\ket{0}+a_1\ket{1}=\sum_{i\in\{0,1\}}a_i\ket{i}$, where each $a_i\in\mathbb{C}$. When a qubit is measured, the wavefunction collapses and the result of the measurement is state $\ket{0}$ with probability $|a_0|^2$ and state $\ket{1}$ with probability $|a_1|^2$ \footnote{We require $|a_0|^2+|a_1|^2=1$}. The space of all possible states of the qubit is called the Hilbert space $\mathcal{H}_1$; the states $\ket{0}$ and $\ket{1}$ provide a basis for this Hilbert space. \footnote{Formally, a Hilbert space is an inner product vector space that is also a complete metric space with respect to the distance function induced by that inner product.}


Multi-qubit systems are represented mathematically by the tensor product of multiple single-qubit systems. Notationally, we write
$$
\ket{\psi_N} = \sum_{\{i_1, i_2,\ldots i_N\}\in\{0,1\}^N}a_{i_1, i_2,\ldots i_N}\ket{i_1, i_2,\ldots i_N}
$$

where the states $\ket{i_1, i_2,\ldots i_N}$ provide a basis for the multi-qubit Hilbert space $\mathcal{H}_N$ and $\ket{\psi_N}$ is generally a superposition of these basis states. A two-qubit state $\ket{\psi}\in \mathcal{H}_2$, cannot necessarily be factorized into two single-qubit states $\ket{\psi_1},\ket{\psi_2}\in\mathcal{H}_1$: 

$$
\ket{\psi} = \sum_{\{i_1,i_2\}\in\{0,1\}^2} a_{i_1,i_2}\ket{i_1,i_2}\neq \ket{\psi_1}\otimes\ket{\psi_2} = \left(a_0\ket{0}+a_1\ket{1}\right)\otimes\left(a_0'\ket{0}+a_1'\ket{1}\right)
$$

we call a state which cannot be so factored a mixed state. In particular, we notice that $\mathcal{H}_2>\mathcal{H}_1\otimes\mathcal{H}_1$, and a similar result holds for multi-qubit systems with $N > 2$.

Under the laws of quantum mechanics, these wavefunctions -- or states -- evolve in time as determined by linear unitary transformations. Furthermore, any operation that is physically possible to perform on a set of qubits can be represented as a unitary operator. We refer to such unitary operators as quantum gates and note that unitary operators can be viewed as rotations in Hilbert space.

A remarkable property of quantum states is their ability to be entangled. Informally, entanglement refers to the property of quantum mechanical systems whereby the state of one qubit cannot be described independently of the other qubits' states. For example, the state $\ket{00}$ is maximally entangled as knowledge of one qubit's state complete specifies the other's. 

In general, quantum algorithms are procedures whereby an initial wavefunction is transformed under a sequence of unitary operations, or quantum gates, and a measurement is made of the transformed state; this measurement is often performed on a readout qubit and is the output of the algorithm. Many quantum computation algorithms are designed to exploit properties of entangled systems (\cite{BernsteinVazirani, DeutschJosza, Simon, Shor, Grover}). 

\subsection{Common Quantum Gates}
\label{subsection:commonquantumgates}

Several common quantum gates are defined below. These gates are defined by their action on the basis states of the Hilbert space, since they extend linearly to superpositions of the basis states. The definitions for the single-qubit Hadamard gate $H$ and the Pauli-$X$ gate $X$ are:
\begin{alignat}{2}
&H\ket{0} = \ket{+} \coloneqq \tfrac{1}{\sqrt{2}}\left(\ket{0}+\ket{1}\right) \qquad &&H\ket{1} = \ket{-} \coloneqq \tfrac{1}{\sqrt{2}}\left(\ket{0}-\ket{1}\right) \\
&X\ket{0} = \ket{1} &&X\ket{1} = \ket{0}
\end{alignat}

A common 2-qubit gate is $CNOT$, which flips the second qubit if the first qubit is in state $\ket{1}$:
\begin{alignat}{2}
&CNOT\ket{00} = \ket{00}\qquad & CNOT\ket{01} = \ket{01} \\ &CNOT\ket{10} = \ket{11} & CNOT\ket{11} = \ket{10} \nonumber
\end{alignat}

\section{Related Work}
\label{section:relatedwork}

Many studies use QNNs to model either inherently quantum or quantum-encoded classical data but are generally restricted to very small images (\cite{Li_2020, Henderson_2020, 9333906}). One line of work encodes classical data in quantum systems and focuses on learning the classifier's circuit architecture. These approaches require an injective map from the input image to a corresponding pure quantum state, which forgoes the exponential compression advantages afforded by methods such as FRQI\footnote{For a demo of a standard approach following \cite{Farhi} using TFQ, which could also serve as a preliminary for this work, see https://www.tensorflow.org/quantum/tutorials/mnist} (\cite{Farhi, CQ1, CQ2, CQ3, CQ4}). Amongst this line of work, \cite{Farhi} propose the general setup that we follow in this paper. In contrast with their work, however, we use the FRQI technique to exploit the dimensionality of the multi-qubit Hilbert space and need much fewer qubits.



Other studies take the quantum wavefunction as given, either by assuming the classical data is already provided in its quantum-encoded form \cite{Schuld} or because they use inherently quantum data (\cite{QQ1, QQ2, QQ3, QQ4, QQ5, QQ6, QQ7, QQ8, QNN3, Caro_2020}). Amongst these papers, \cite{Schuld} is perhaps closest to this work. The authors, however, take the mixed-state encodings of images as given for input to a QNN and do not describe how to construct the quantum states.
Other work, such as \cite{QNN3}, assumes the wavefunctions as given proposes a generalization of the perceptron to the quantum setting, which provides a more generalized framework than in \cite{Farhi} 
These authors use inherently quantum data, in contrast with our work. We use classical data and explicitly construct quantum circuits to encode \textit{classical} images into their wavefunctions. Our approach lends itself to direct experimentation and is usable with modern quantum machine learning packages.

Finally, a third line of work uses quantum convolutional neural networks via semi-classical simulations meant to model the noise introduced by quantum effects, as in \cite{QCNN}. These approaches do not provide a fully quantum simulation to evolve the quantum states, which would require construction of the actual data's wavefunctions as in our work. 

Throughout prior work, encoding classical data in quantum states \textit{efficiently} appears to be a common open problem.


\section{Formal Setting}
\label{section:formalsetting}

\subsection{Problem Statement}
\label{subsection:problemstatement}

In classical image classification, the input to our model is an $n\times n$-pixel image $\in\{0,1\}^{2n}$ and our goal is to learn a classification function with binary outputs $f_{\text{classical}}$ parametrized by weights $w$:

\begin{align}
f_{\text{classical}}(w) : \{0,1\}^{2n} \rightarrow \{0,1\}
\end{align}

In the quantum setting, the input to our classification function is still an $n\times n$-pixel image but must be encoded in a $\lceil \text{log}2n \rceil + 1$ dimensional Hilbert space $\mathcal{H}$ by an encoding function $\mathcal{F}$, where the $+1$ is for the readout qubit. The quantum neural network is a sequence of unitary operations $\mathcal{U} = U_1 \circ U_{2} \circ \dots U_N$ parametrized by angles $\theta = \theta_1, \theta_2, \dots ,\theta_N$. To obtain a classification prediction, a measurement is performed on the readout qubit:

\begin{align}
    f_{\text{quantum}}(\theta)    &: \{0,1\}^{2n} \rightarrow \{0,1\} \\
                    &: \{0,1\}^{2n} \xrightarrow[]{\,\,\mathcal{F}\,\,} \mathcal{H} 
                                 \xrightarrow[]{\,\,\mathcal{U}(\theta)\,\,} \mathcal{H}\xrightarrow[]{\text{measure}}\{0,1\} \nonumber
\end{align}

That is $f_\text{quantum}(\theta) = \text{measure}\circ \mathcal{U}(\theta)\circ \mathcal{F}$. 

In Sections \ref{section:methods} and \ref{section:methods} we propose an implementation of the FRQI algorithm \cite{FRQI} to construct $\mathcal{F}$, propose a construction of $U(\theta)$, describe how to learn the parameters $\theta$ via standard backpropagation, and describe the final measurement step.

\subsection{Dataset and Quantum Encoding}
\label{subsection:quantumencoding}

Crucial to our approach is the encoding of a classical datapoint (e.g. an image) in a quantum state. In our experiments, we use the MNIST dataset of handwritten digits \cite{MNIST}. Following \cite{Farhi}, we restrict our dataset to those of only two ground truth labels: $3$ and $6$. We downsample image resolutions to either $8\times8$ or $16\times16$ using bilinear interpolation. The remaining dataset is approximately $12,000$ training images and $1,100$ validation images for each resolution. Finally, we transform the images to black and white by thresholding the pixel color.

In Figure \ref{fig:digit}, we present an MNIST image downsampled to different resolutions. Prior work uses resolutions of only $4\times4$, but loses many important features of the original data \cite{Farhi}. With the FRQI encoding and the further compression we are able to encode higher-resolution images on current quantum hardware; this insight motivates our study of different downsampled resolutions.

{
\setlength{\belowcaptionskip}{-10pt}
\begin{figure}
\centering
\includegraphics[width=0.8\textwidth,keepaspectratio]{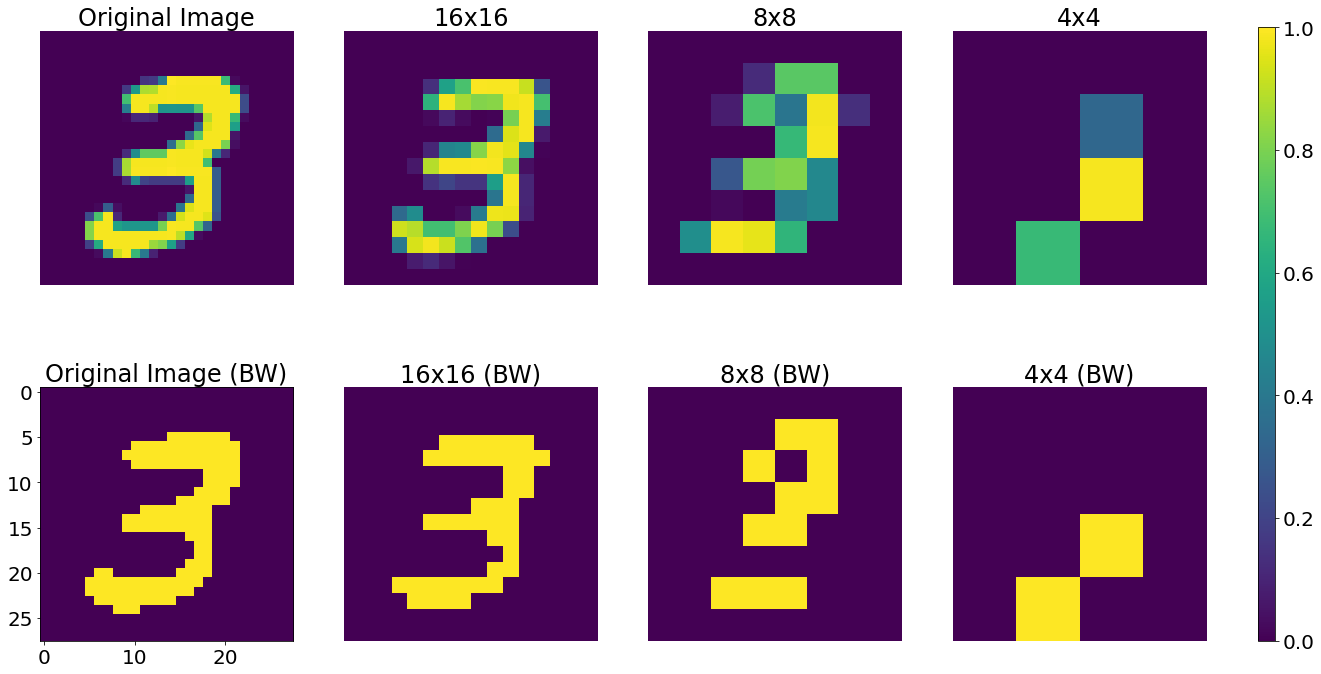}
\caption{A digit from MNIST presented at different downsampled resolutions (downscaled resolutions indicated on top of each image. The top row consists of grayscale images with $0\leq\text{color}\leq1$ as in the original dataset. The bottom row are black and white images obtained by thresholding the pixel color from the respective images above.}
\label{fig:digit}
\end{figure}
}

After preprocessing, each image is a black and white $2^n\times 2^n = 2^{2n}$ dimensional binary vector. Our objective is to encode the image as a wavefunction $\ket{\psi_\mathrm{data}}$:

\begin{equation}
\label{eqn:wavefn}
\ket{\psi_\mathrm{data}} = \sum_{q:= \{q_0,q_1,\ldots,q_{2n-1}\}\in\{0,1\}^{2n}} \ket{q_0,q_1,\ldots,q_{2n-1}}\otimes \left( \cos \theta_q \ket{0} + \sin \theta_q \ket{1}\right)
\end{equation}

In Equation \ref{eqn:wavefn}, each basis state $\ket{q_0,q_1,\ldots,q_{2n-1}}$ of the "pixel qubits" represents a possible bitstring of length $2^{2n}$ with the strength of the superposition component and color determined by $\theta_q$ in the "color qubit". In our dataset, each $\theta_q$ is either $0$ or $\frac{\pi}{2}$.

In some experiments in Section \ref{section:methods}, we also consider allocating more qubits to encode the color angle $\theta_q$ instead of the pixel locations:

$$
\ket{\psi_\mathrm{data}} = \sum_{q:= \{q_0,q_1,\ldots,q_{2n-1}\}\in\{0,1\}^{2n}} \ket{q_0,q_1,\ldots,q_{2n-3}}\otimes \left( \cos \tilde{\theta}_q \ket{0} + \sin \tilde{\theta}_q \ket{1}\right)
$$

To do this, we exploit the observation that the color qubit is always either $\ket{0}$ or $\ket{1}$ and map into it the last two pixel qubits according to the transformation: 

\begin{equation}
\ket{q_{2n-2}, q_{2n-1}}\otimes\ket{q_c}\rightarrow\ket{\tilde{q}_c}=\cos \tilde{\theta}_q \ket{0} + \sin \tilde{\theta}_q \ket{1}
\end{equation}

where each $q_i\in\{0,1\}$ and

\begin{equation}
\tilde{\theta}_q=\frac{\pi}{2}\left(q_c+ \frac{q_{2n-2}}{2} + \frac{q_{2n-1}}{4}\right)=\theta_q + \frac{\pi}{4}\left(q_{2n-2} + \frac{q_{2n-1}}{2}\right) \label{eq:compression}
\end{equation}

\section{Methods}
\label{section:methods}

\subsection{Encoding the Images in Wavefunctions}
\label{subsection:encodingmethods}

\begin{figure}[H]
\centering
\includegraphics[width=\textwidth,keepaspectratio]{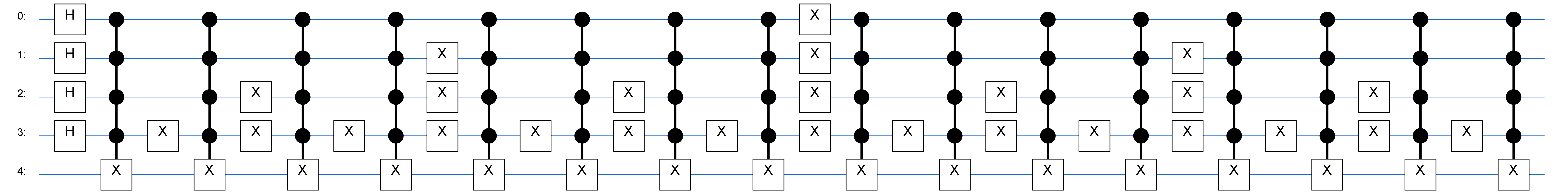}
\caption{An example circuit to construct the superposition state for $4$ qubits, using $4$-qubit generalized TOFFOLI gate. We follow standard quantum circuit diagram conventions in which the dots represent the control of a gate by the given qubits.}
\label{fig:circuit}
\end{figure}

In our approach, we must first encode the image in a quantum wavefunction. We pass an initial state of $\ket{0\dots0}$ through a quantum circuit with a given structure, demonstrated by Figure \ref{fig:circuit} for a $4$-qubit state. Initially, a Hadamard operation $H^{\otimes 2n}$ is performed on the $2n$ pixel qubits. This is followed by a series of $n$-controlled $X$-gates (also known as generalized TOFFOLI gates \cite{Toffoli, NToffoli1, NToffoli2}) with alternating $X$ gates that determine the color qubits which will be transformed. The $n$-qubit circuit is constructed recursively from smaller-qubit circuits by observing the symmetries in the construction. 


To construct this circuit we need to define the generalized TOFFOLI gate for $n$ qubits from basic two-qubit gates; this also enables our implementation in standard packages such as Cirq (\cite{Cirq}) that only support backpropagation through two-qubit gates. We use the following lemma, first shown in \cite{NControl}, to recursively decompose the $n$-qubit generalized TOFFOLI gate as a sequence of $(n-1)$-qubit generalized TOFFOLI gates and CNOT gates:

\begin{lemma}
\label{lemma:gatedecomp}
(\cite{NControl}, Lemma 7.5): For a rotation matrix $R(t)$, an $n$-controlled rotation gate can be decomposed into a circuit of the form shown in Figure \ref{fig:lemma1}.
\begin{figure}[H]
\begin{equation}
\includegraphics[height=4cm,valign=c]{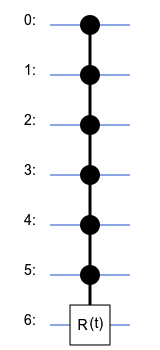}
=
\includegraphics[height=4cm,valign=c]{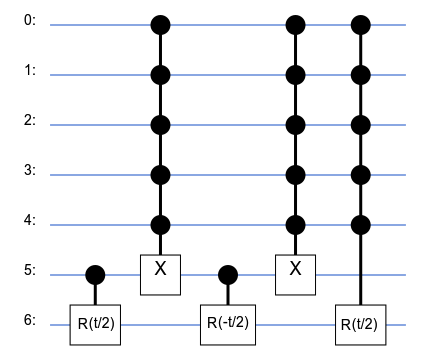}
\end{equation}
\caption{$n-$qubit controlled gate in terms of $(n-1)-$qubit controlled gates}
\label{fig:lemma1}
\end{figure}
\end{lemma}

From the recursive properties of the diagram above we see that an $n$-qubit controlled rotation decomposes into $\left(2\cdot 3^n - 1\right)$ one-qubit controlled rotations. Since we flip at most $2^n$ pixels for a given image, we see that for each image
$$
\text{\# one-qubit controlled gates }\leq 2^n \left(2\cdot 3^n - 1\right)
$$ 
in addition to the standalone $X-$gates to encode the $2^{n}\times 2^{n}$ image in the amplitudes of the input wavefunction. Though this bound is exponential in $n$, we find this acceptable as it is still classically simulable for larger images that are primarily constrained by the size of their qubit representations.


\subsection{Readout Qubit and Predicted Labels}
\label{subsection:readoutqubit}

Our wavefunction must also contain a readout qubit on which we perform measurements that will be the model's predicted labels. As such, we prepare the wavefunction $\ket{\psi_\mathrm{in}}=\ket{\psi_\mathrm{data}}\otimes\ket{\mathrm{readout}}$. We choose the $Z$-gate for measurement and thus initialize the readout qubit in the $\ket{+}$ state, which is common practice to produce an initially unbiased output:

\begin{equation}
\ket{\psi_\mathrm{in}}=\ket{\psi_\mathrm{data}}\otimes H\ket{0}=\ket{\psi_\mathrm{data}}\otimes\ket{+}
\end{equation}

where $\ket{\psi_\mathrm{data}}$ is prepared as in Section \ref{subsection:encodingmethods}. 

The model which is used to transform $\ket{\psi_\mathrm{in}}$ is a QNN with $L$ layers. Following \cite{Farhi}, each layer is represented by a  parametrized unitary matrix. The model's output state is:

\begin{equation}
\ket{\psi_\mathrm{out}(\theta)} = U(\theta_L)\ldots U(\theta_1)\left( \ket{\psi_\mathrm{data}}\otimes\ket{+}\right)
\end{equation}

where $\theta \coloneqq (\theta_1, \ldots , \theta_L)$. The final measurement is performed with $Z-$gate on the readout qubit; the predicted label is $\bra{\psi_\mathrm{out}(\theta)} \mathbb{I}^{2n+1}\otimes Z \ket{\psi_\mathrm{out}(\theta)}$. We train the model's parameters, $\theta_1 \ldots \theta_L$, via stochastic gradient descent (SGD) using these predictions and the hinge loss:

\begin{equation}
\mathrm{loss}^{(i)}(\theta) = 1 - y^{(i)}\bra{\psi_\mathrm{out}^{(i)}(\theta)} \mathbb{I}^{2n+1}\otimes Z \ket{\psi_\mathrm{out}^{(i)}(\theta)}
\end{equation}

where the superscript $(i)$ is used to refer to the $i^{th}$ training example.

\subsection{Implementation Details}
\label{subsection:implementationdetails}

We use Cirq (\cite{Cirq}) to encode the images into their respective wavefunctions and TensorFlow Quantum (TFQ) (\cite{TFQ}) to train the model via the paradigm described in Section \ref{subsection:readoutqubit}. TFQ permits the use of Parametrized Quantum Circuits (PQCs), which describe the unitary operations of the QNN, as a single Keras layer \cite{chollet2015keras} within the standard TensorFlow framework. 


Backpropagation through quantum layers is nontrivial. We recall that, for any layer, the $2^n\times2^n$ unitary operator can be expressed in terms of the exponential of a $2^n\times2^n$ Hermitian operator $H$ called the Hamiltonian, which in turn can be decomposed into its Pauli decomposition (a tensor product of $n$ Pauli matrices, which form an orthonormal basis over the Hilbert space of Hermitian matrices over $\mathbb{R}$):

\begin{equation}\label{eqn:PauliDecomposition}
U(\theta) = \exp\left\{ i\sum_{\sigma_i\in 
\{\mathbb{I},\sigma_1,\sigma_2,\sigma_3\}
^n}\theta^{(\sigma_i)}\bigotimes_{i} \sigma_i\right\}
\end{equation}

When the layer is restricted to unitary operations whose Hamiltonian has a single term in the Pauli decomposition, 
gradients with respect to the layer's parameters can be computed analytically using the parameter shift techniques first introduced in \cite{Grad0} and \cite{Grad1} (\cite{Grad2}). These techniques provide a way to calculate partial derivatives of parameterized quantum circuits in terms of other functions that use the same circuit architecture with shifted parameters. 
For this reason, we restrict the gates in our quantum layers to be multi-qubit exponential Pauli gates. For these gates, analytic gradients can also be computed because they are rotations of operations whose Hamiltonians contain a single type of term. For example, the $XX$-gate can be written:

$$
\left(X\otimes X\right)^\theta =  \exp \left\{ \theta\left(-i\tfrac{\pi}{2} \left(X-\mathbb{I}\right) \otimes -i\tfrac{\pi}{2} \left(X-\mathbb{I}\right)\right)\right\} =  e^{ -i\tfrac{\pi}{2}\theta \left(X-\mathbb{I}\right)}\otimes  e^{ -i\tfrac{\pi}{2}\theta \left(X-\mathbb{I}\right)}
$$

\subsection{Network Architecture}

A general $2^n\times 2^n$ learnable unitary operation would consist of $2^{2n}$ trainable real parameters and the standard representation of this parameters follows equation \eqref{eqn:PauliDecomposition}. Note that, as described in Section \ref{subsection:implementationdetails}, this does not necessarily permit analytic gradient computation in the backpropagation step.

To permit analytical gradient computations, we construct layers having a specific structure; an example such layer is demonstrated in Figure \ref{fig:layerCRADL}. Each layer consists of either $XX$ or $ZZ$ operators applied in succession to each pixel qubit and the readout qubit, followed by the same operation on the pixel qubit and the color qubit. Empirically, the Color-Readout-Alternating-Double-Layer architecture (CRADL) presented in Figure \ref{fig:layerCRADL} resulted in the best performance.

Each consecutive pixel-readout and pixel-color pair of gates share the same learning parameter, a rotation angle. One double layer of the type shown in Figure \ref{fig:layerCRADL} will have $2n$ learnable parameters, where $n$ is the number of pixel qubits. A circuit with $L$ layers will therefore have $nL$ parameters. In the experiments in Section \ref{section:results}, we choose the number of layers such that the number of parameters $nL$ is comparable to those of the classical benchmark, given a fixed number of qubits.

{
\setlength{\belowcaptionskip}{-10pt}
\begin{figure}[H]
\centering
\includegraphics[width=1.0\textwidth,keepaspectratio]{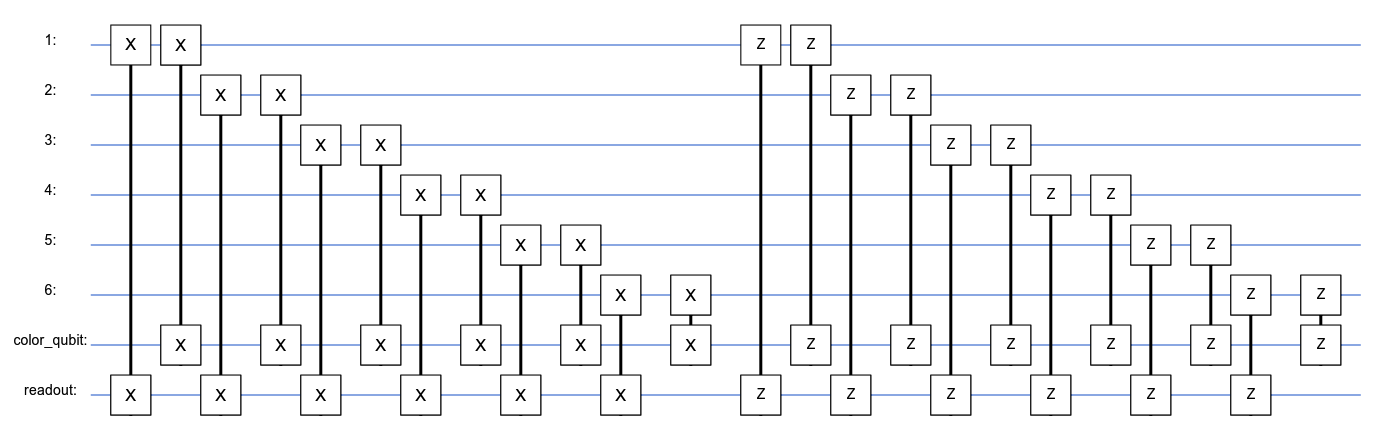}
\caption{\small A ``CRADL'' network double layer with $6$ pixel qubits, consisting of consecutive pixel-readout pixel-color $XX$ gates, followed by analogous $ZZ$ gates}
\label{fig:layerCRADL}
\end{figure}
}

We note that there are equivalent network architectures that lead to comparable results, such as the Color-Readout-Alternating-Mixed-Layer architecture shown in Figure \ref{fig:layerCRAML}.

{
\setlength{\belowcaptionskip}{-10pt}
\begin{figure}[H]
\centering
\includegraphics[width=1.0\textwidth,keepaspectratio]{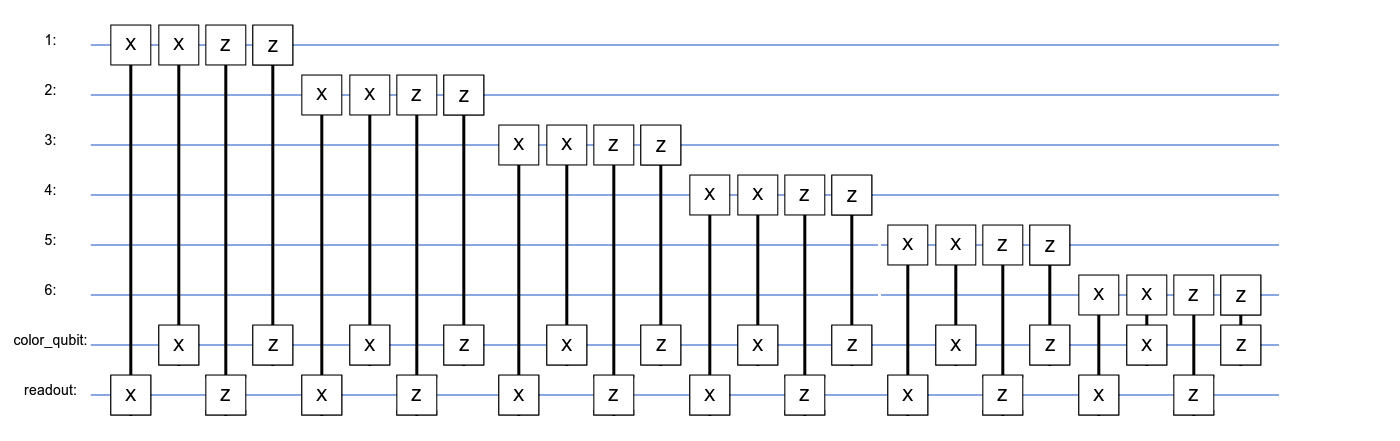}
\caption{\small A ``CRAML'' network layer with $6$ pixel qubits, consisting of consecutive pixel-readout pixel-color $XX$ and $ZZ$ gates}
\label{fig:layerCRAML}
\end{figure}
}

\section{Results}
\label{section:results}

We benchmark our quantum learning framework against a classical neural network with two hidden layers with ReLU activations and a single-neuron output layer; in this setting, the quantum and classical models have a comparable number of parameters. We trained the quantum neural network for $10$ epochs, which is the same number of epochs after which the classical neural network began to overfit (as determined by cross-validation). 
All experiments were conducted on a personal laptop with no GPU (Macbook Pro, 2.4 GHz 8-Core Intel Core i9 CPU, 64 GB 2667 MHz DDR4 RAM).



\begin{table}[H]

\setlength{\tabcolsep}{16pt}
\begin{center}
\begin{tabular}{lll}
\multicolumn{1}{c}{\bf Network}  &\multicolumn{1}{c}{\bf $8\times8$ Image}
&\multicolumn{1}{c}{\bf $16\times16$ Image}
\\ \hline \\
Classical CNN          &$94\pm1\%$     & $98.9\pm0.3\%$\\
Quantum CRADL          &$92\pm1\%$     & $-$\\
Quantum CRADL $-2Q$   &$88\pm1\%$     & $90\pm1\%$\\
\end{tabular}
\end{center}
\caption{Test accuracies after the 10th training epoch for classical and quantum networks.}
\label{table:ClassicalQuantumComparison}
\end{table}

{
\setlength{\belowcaptionskip}{-10pt}
\begin{figure}[H]
\centering
\includegraphics[width=\textwidth,keepaspectratio]{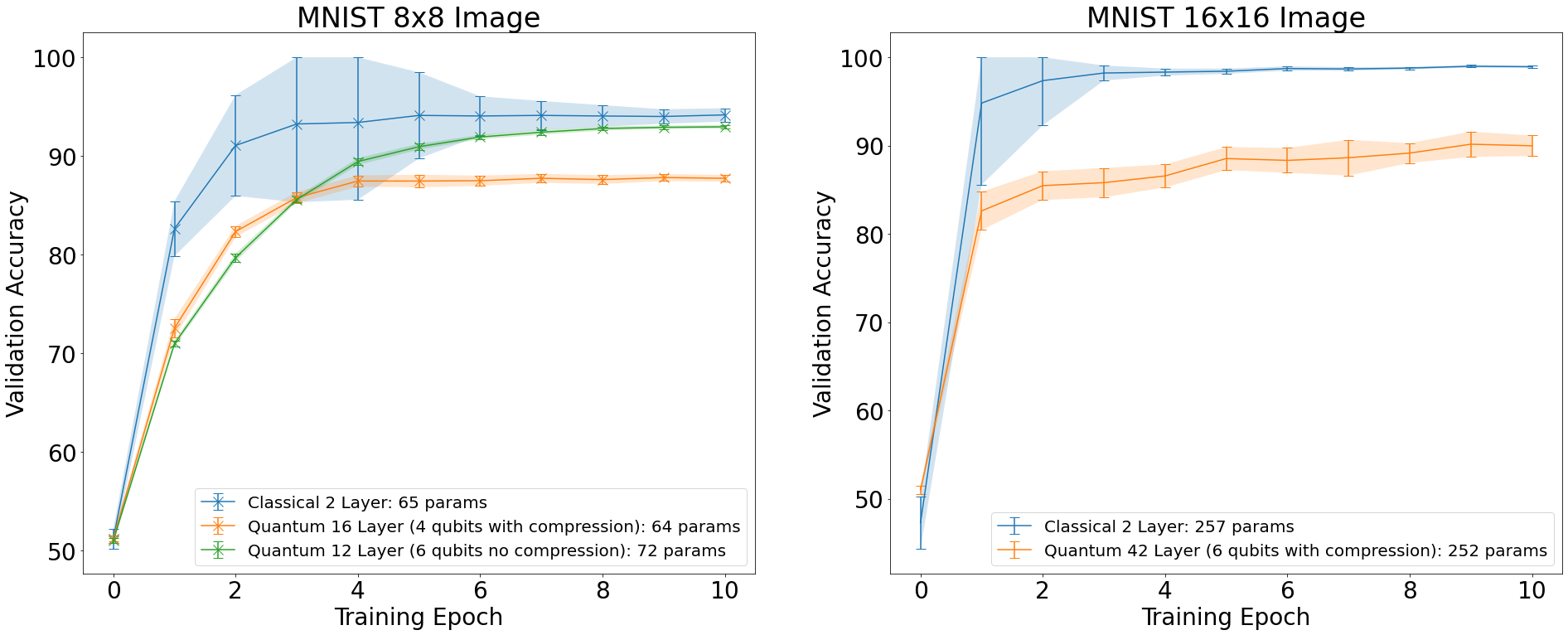}
\caption{\small Test accuracy versus training epoch for classical and quantum models, with and without the extra 2 qubit compression described in Section \ref{subsection:quantumencoding}. When images are embedded on 6 qubits (green curve, left), we achieve performance comparable to classical networks with the same number of parameters. When images are further compressed (orange curves), performance degrades.} 

\label{fig:experiment}
\end{figure}
}

Table \ref{table:ClassicalQuantumComparison} demonstrates our final results. On $8\times 8$ images, the QNN without the extra two-qubit compression achieves performance comparable to the classical network, whereas the network with the extra two-qubit compression (denoted $-2Q$) performs worse. For $16\times16$ images, we add more layers to the QNN so that the the number of parameters remains comparable to the classical dense network and must use the extra two-qubit compression due to the computational cost of the experiment. We observe that the QNN is unable to achieve the same performance as the classical network, likely due to the extensive compression of the images in the quantum states.

Figure \ref{fig:experiment} shows the validation accuracy of our quantum and baseline classical models versus training epoch. We observe that the classical neural network and both quantum neural networks demonstrate similar validation performance curves.

\section{Discussion and Conclusion}
\label{section:conclusion}

We note that the method we describe in Equation \ref{eq:compression} to reduce the number of required qubits results in worse performance. This degradation in test accuracy may be palatable in applications that attempt to minimize their qubit usage. When we attempted to lower the number of necessary qubits further, we observed unstable learning behavior. In such settings with reduced feature dimensionality, it may be necessary to redesign the network architecture.


In this paper, we developed a proof of concept for recently proposed QNN models. In the process, we proposed a methodology to map classical images to quantum states that may be of independent interest to the community. We also propose a new form of quantum neural network layers motivated by the highly entangled input states, the CRADL and CRAML layers in Section \ref{subsection:implementationdetails}, and demonstrate that a model consisting of these layers achieves performance competitive with classical neural networks with a comparable number of parameters. Furthermore, our work is evidence that quantum machine learning algorithms can scale to data of dimensions larger than those previously tractable by classical simulation or available quantum hardware and classify MNIST images of size $16 \times 16$ on a personal laptop.

\section{Future Work}
\label{section:futurework}

We did not do a comprehensive survey of the space of all possibly unitary operations that could be used for each hidden layer, but we could imagine that invoking $3$ or more qubit gates to the layer circuits would improve the learning outcomes, as that would give more direct access to the correlation structure. The circuit above Figure \ref{fig:layerCRADL} composed entirely of two-qubit gates is trying to work around this limitation.

We would like in the future to conduct a more systematic survey of network architectures to enhance the learning outcomes with possibly more involved quantum gates, either within TensorFlow Quantum or implemented separately, and study the cost benefit of forgoing analytic gradients for some of these gates for more flexibility in construction. Another lower hanging fruit is optimizing the encoded input circuits in terms of memory. As long as we write and simulate quantum algorithms on digital computers, representing the wavefunction in this manner seems inevitable and we should find better optimizations to describe the encoding procedure through circuits with much less gates than the nice recursive algorithms discussed in this paper.

We observed in Section \ref{section:results} that compressing the images into larger quantum systems resulted in better performance at the expensive of greater complexity in physical realization. We leave a more thorough analysis of this tradeoff, including the understanding of the interplay between the data qubits and the color qubit, to future work.

We also note that we may view the encoding with limited circuit gates or on limited qubits as a form of implicit regularization, which has been observed to improve generalization performance as in \cite{implicitreg}. We leave investigation of these connections to future work.


\section*{Ethics Statement}
\label{section:ethics}

We do not foresee any ethical concerns with our work.

\section*{Reproducibility Statement}
\label{section:reproducibility}

All code needed to reproduce the experiments in this paper can be found in the following repository \url{https://github.com/ThrunGroup/quantum_ml}. The code contains a \texttt{README.md} file with instructions on how to reproduce the experimental results. The only dataset required is the MNIST dataset, which is easily obtained from \url{http://yann.lecun.com/exdb/mnist/}. The preprocessing applied to the MNIST dataset is described in detail in Section \ref{section:methods}.


\bibliographystyle{iclr2022_conference}

\begin{thebibliography}{46}
\providecommand{\natexlab}[1]{#1}
\providecommand{\url}[1]{\texttt{#1}}
\expandafter\ifx\csname urlstyle\endcsname\relax
  \providecommand{\doi}[1]{doi: #1}\else
  \providecommand{\doi}{doi: \begingroup \urlstyle{rm}\Url}\fi

\bibitem[A{\"i}meur et~al.(2013)A{\"i}meur, Brassard, and Gambs]{CQ1}
Esma A{\"i}meur, Gilles Brassard, and S{\'e}bastien Gambs.
\newblock Quantum speed-up for unsupervised learning.
\newblock \emph{Machine Learning}, 90\penalty0 (2):\penalty0 261--287, Feb
  2013.
\newblock ISSN 1573-0565.
\newblock \doi{10.1007/s10994-012-5316-5}.
\newblock URL \url{https://doi.org/10.1007/s10994-012-5316-5}.

\bibitem[Alvarez-Rodriguez et~al.(2017)Alvarez-Rodriguez, Lamata,
  Escandell-Montero, Mart{\'i}n-Guerrero, and Solano]{QQ6}
Unai Alvarez-Rodriguez, Lucas Lamata, Pablo Escandell-Montero, Jos{\'e}~D.
  Mart{\'i}n-Guerrero, and Enrique Solano.
\newblock Supervised quantum learning without measurements.
\newblock \emph{Scientific Reports}, 7\penalty0 (1):\penalty0 13645, Oct 2017.
\newblock ISSN 2045-2322.
\newblock \doi{10.1038/s41598-017-13378-0}.
\newblock URL \url{https://doi.org/10.1038/s41598-017-13378-0}.

\bibitem[Arute et~al.(2019)Arute, Arya, Babbush, Bacon, Bardin, Barends,
  Biswas, Boixo, Brandao, Buell, Burkett, Chen, Chen, Chiaro, Collins,
  Courtney, Dunsworth, Farhi, Foxen, Fowler, Gidney, Giustina, Graff, Guerin,
  Habegger, Harrigan, Hartmann, Ho, Hoffmann, Huang, Humble, Isakov, Jeffrey,
  Jiang, Kafri, Kechedzhi, Kelly, Klimov, Knysh, Korotkov, Kostritsa, Landhuis,
  Lindmark, Lucero, Lyakh, Mandr{\`a}, McClean, McEwen, Megrant, Mi,
  Michielsen, Mohseni, Mutus, Naaman, Neeley, Neill, Niu, Ostby, Petukhov,
  Platt, Quintana, Rieffel, Roushan, Rubin, Sank, Satzinger, Smelyanskiy, Sung,
  Trevithick, Vainsencher, Villalonga, White, Yao, Yeh, Zalcman, Neven, and
  Martinis]{Sycamore1}
Frank Arute, Kunal Arya, Ryan Babbush, Dave Bacon, Joseph~C. Bardin, Rami
  Barends, Rupak Biswas, Sergio Boixo, Fernando G. S.~L. Brandao, David~A.
  Buell, Brian Burkett, Yu~Chen, Zijun Chen, Ben Chiaro, Roberto Collins,
  William Courtney, Andrew Dunsworth, Edward Farhi, Brooks Foxen, Austin
  Fowler, Craig Gidney, Marissa Giustina, Rob Graff, Keith Guerin, Steve
  Habegger, Matthew~P. Harrigan, Michael~J. Hartmann, Alan Ho, Markus Hoffmann,
  Trent Huang, Travis~S. Humble, Sergei~V. Isakov, Evan Jeffrey, Zhang Jiang,
  Dvir Kafri, Kostyantyn Kechedzhi, Julian Kelly, Paul~V. Klimov, Sergey Knysh,
  Alexander Korotkov, Fedor Kostritsa, David Landhuis, Mike Lindmark, Erik
  Lucero, Dmitry Lyakh, Salvatore Mandr{\`a}, Jarrod~R. McClean, Matthew
  McEwen, Anthony Megrant, Xiao Mi, Kristel Michielsen, Masoud Mohseni, Josh
  Mutus, Ofer Naaman, Matthew Neeley, Charles Neill, Murphy~Yuezhen Niu, Eric
  Ostby, Andre Petukhov, John~C. Platt, Chris Quintana, Eleanor~G. Rieffel,
  Pedram Roushan, Nicholas~C. Rubin, Daniel Sank, Kevin~J. Satzinger, Vadim
  Smelyanskiy, Kevin~J. Sung, Matthew~D. Trevithick, Amit Vainsencher, Benjamin
  Villalonga, Theodore White, Z.~Jamie Yao, Ping Yeh, Adam Zalcman, Hartmut
  Neven, and John~M. Martinis.
\newblock Quantum supremacy using a programmable superconducting processor.
\newblock \emph{Nature}, 574\penalty0 (7779):\penalty0 505--510, Oct 2019.
\newblock ISSN 1476-4687.
\newblock \doi{10.1038/s41586-019-1666-5}.
\newblock URL \url{https://doi.org/10.1038/s41586-019-1666-5}.

\bibitem[Barenco et~al.(1995)Barenco, Bennett, Cleve, DiVincenzo, Margolus,
  Shor, Sleator, Smolin, and Weinfurter]{NControl}
Adriano Barenco, Charles~H. Bennett, Richard Cleve, David~P. DiVincenzo, Norman
  Margolus, Peter Shor, Tycho Sleator, John~A. Smolin, and Harald Weinfurter.
\newblock Elementary gates for quantum computation.
\newblock \emph{Phys. Rev. A}, 52:\penalty0 3457--3467, Nov 1995.
\newblock \doi{10.1103/PhysRevA.52.3457}.
\newblock URL \url{https://link.aps.org/doi/10.1103/PhysRevA.52.3457}.

\bibitem[Beer et~al.(2020)Beer, Bondarenko, Farrelly, Osborne, Salzmann,
  Scheiermann, and Wolf]{QNN3}
Kerstin Beer, Dmytro Bondarenko, Terry Farrelly, Tobias~J. Osborne, Robert
  Salzmann, Daniel Scheiermann, and Ramona Wolf.
\newblock Training deep quantum neural networks.
\newblock \emph{Nature Communications}, 11\penalty0 (1), Feb 2020.
\newblock ISSN 2041-1723.
\newblock \doi{10.1038/s41467-020-14454-2}.
\newblock URL \url{http://dx.doi.org/10.1038/s41467-020-14454-2}.

\bibitem[Bernstein \& Vazirani(1997)Bernstein and Vazirani]{BernsteinVazirani}
Ethan Bernstein and Umesh Vazirani.
\newblock Quantum complexity theory.
\newblock \emph{SIAM Journal on Computing}, 26\penalty0 (5):\penalty0
  1411--1473, 1997.
\newblock \doi{10.1137/S0097539796300921}.
\newblock URL \url{https://doi.org/10.1137/S0097539796300921}.

\bibitem[Boixo et~al.(2018)Boixo, Isakov, Smelyanskiy, Babbush, Ding, Jiang,
  Bremner, Martinis, and Neven]{Sycamore2}
Sergio Boixo, Sergei~V. Isakov, Vadim~N. Smelyanskiy, Ryan Babbush, Nan Ding,
  Zhang Jiang, Michael~J. Bremner, John~M. Martinis, and Hartmut Neven.
\newblock Characterizing quantum supremacy in near-term devices.
\newblock \emph{Nature Physics}, 14\penalty0 (6):\penalty0 595--600, Jun 2018.
\newblock ISSN 1745-2481.
\newblock \doi{10.1038/s41567-018-0124-x}.
\newblock URL \url{https://doi.org/10.1038/s41567-018-0124-x}.

\bibitem[Caro \& Datta(2020)Caro and Datta]{Caro_2020}
Matthias~C. Caro and Ishaun Datta.
\newblock Pseudo-dimension of quantum circuits.
\newblock \emph{Quantum Machine Intelligence}, 2\penalty0 (2), Nov 2020.
\newblock ISSN 2524-4914.
\newblock \doi{10.1007/s42484-020-00027-5}.
\newblock URL \url{http://dx.doi.org/10.1007/s42484-020-00027-5}.

\bibitem[Chollet et~al.(2015)]{chollet2015keras}
Fran\c{c}ois Chollet et~al.
\newblock Keras.
\newblock \url{https://github.com/fchollet/keras}, 2015.

\bibitem[Cirq(2021)]{Cirq}
Cirq.
\newblock Cirq, August 2021.
\newblock URL \url{https://doi.org/10.5281/zenodo.5182845}.
\newblock {See full list of authors on Github:
  https://github.com/quantumlib/Cirq/graphs/contributors}.

\bibitem[David \& Richard(1992)David and Richard]{DeutschJosza}
Deutsch David and Jozsa Richard.
\newblock Rapid solution of problems by quantum computation.
\newblock \emph{Proc. R. Soc. Lond. A}, 439:\penalty0 553–558, 1992.
\newblock URL \url{http://doi.org/10.1098/rspa.1992.0167}.

\bibitem[Du et~al.(2020)Du, Hsieh, Liu, and Tao]{QQ7}
Yuxuan Du, Min-Hsiu Hsieh, Tongliang Liu, and Dacheng Tao.
\newblock Expressive power of parametrized quantum circuits.
\newblock \emph{Phys. Rev. Research}, 2:\penalty0 033125, Jul 2020.
\newblock \doi{10.1103/PhysRevResearch.2.033125}.
\newblock URL \url{https://link.aps.org/doi/10.1103/PhysRevResearch.2.033125}.

\bibitem[Dunjko et~al.(2016)Dunjko, Taylor, and Briegel]{QQ4}
Vedran Dunjko, Jacob~M. Taylor, and Hans~J. Briegel.
\newblock Quantum-enhanced machine learning.
\newblock \emph{Phys. Rev. Lett.}, 117:\penalty0 130501, Sep 2016.
\newblock \doi{10.1103/PhysRevLett.117.130501}.
\newblock URL \url{https://link.aps.org/doi/10.1103/PhysRevLett.117.130501}.

\bibitem[E.~Farhi(2018)]{Farhi}
H.~Neven E.~Farhi.
\newblock {Classification with Quantum Neural Networks on Near Term
  Processors}.
\newblock \emph{Arxiv Preprint}, 2018.
\newblock URL \url{https://arxiv.org/abs/1802.06002}.

\bibitem[Gambs(2008)]{QQ2}
Sébastien Gambs.
\newblock Quantum classification, 2008.

\bibitem[Gi(2016)]{DeepLearning}
Kim~Kwang Gi.
\newblock Book review: Deep learning.
\newblock \emph{Healthc Inform Res}, 22\penalty0 (4):\penalty0 351--354, 2016.
\newblock \doi{10.4258/hir.2016.22.4.351}.
\newblock URL \url{http://e-hir.org/journal/view.php?number=828}.

\bibitem[GoogleAI(2020)]{TFQ}
GoogleAI.
\newblock Tensorflow quantum: A software framework for quantum machine
  learning.
\newblock \emph{Arxiv Preprint}, 2020.
\newblock URL \url{https://arxiv.org/abs/2003.02989}.

\bibitem[Grover(1996)]{Grover}
Lov~K. Grover.
\newblock A fast quantum mechanical algorithm for database search.
\newblock In \emph{Proceedings of the Twenty-Eighth Annual ACM Symposium on
  Theory of Computing}, STOC '96, pp.\  212–219, New York, NY, USA, 1996.
  Association for Computing Machinery.
\newblock ISBN 0897917855.
\newblock \doi{10.1145/237814.237866}.
\newblock URL \url{https://doi.org/10.1145/237814.237866}.

\bibitem[Harrow \& Napp(2021)Harrow and Napp]{Grad2}
Aram~W. Harrow and John~C. Napp.
\newblock Low-depth gradient measurements can improve convergence in
  variational hybrid quantum-classical algorithms.
\newblock \emph{Physical Review Letters}, 126\penalty0 (14), Apr 2021.
\newblock ISSN 1079-7114.
\newblock \doi{10.1103/physrevlett.126.140502}.
\newblock URL \url{http://dx.doi.org/10.1103/PhysRevLett.126.140502}.

\bibitem[Henderson et~al.(2020)Henderson, Shakya, Pradhan, and
  Cook]{Henderson_2020}
Maxwell Henderson, Samriddhi Shakya, Shashindra Pradhan, and Tristan Cook.
\newblock Quanvolutional neural networks: powering image recognition with
  quantum circuits.
\newblock \emph{Quantum Machine Intelligence}, 2\penalty0 (2), feb 2020.
\newblock \doi{10.1007/s42484-020-00012-y}.
\newblock URL \url{https://doi.org/10.1007/s42484-020-00012-y}.

\bibitem[Kapoor et~al.(2016)Kapoor, Wiebe, and Svore]{CQ4}
Ashish Kapoor, Nathan Wiebe, and Krysta Svore.
\newblock Quantum perceptron models.
\newblock In D.~Lee, M.~Sugiyama, U.~Luxburg, I.~Guyon, and R.~Garnett (eds.),
  \emph{Advances in Neural Information Processing Systems}, volume~29. Curran
  Associates, Inc., 2016.
\newblock URL
  \url{https://proceedings.neurips.cc/paper/2016/file/d47268e9db2e9aa3827bba3afb7ff94a-Paper.pdf}.

\bibitem[Kerenidis et~al.(2019)Kerenidis, Landman, and Prakash]{QCNN}
Iordanis Kerenidis, Jonas Landman, and Anupam Prakash.
\newblock Quantum algorithms for deep convolutional neural networks, 2019.

\bibitem[LeCun \& Cortes(2010)LeCun and Cortes]{MNIST}
Yann LeCun and Corinna Cortes.
\newblock {MNIST} handwritten digit database.
\newblock 2010.
\newblock URL \url{http://yann.lecun.com/exdb/mnist/}.

\bibitem[LeCun et~al.(2015)LeCun, Bengio, and Hinton]{LeCun}
Yann LeCun, Yoshua Bengio, and Geoffrey Hinton.
\newblock Deep learning.
\newblock \emph{Nature}, 521\penalty0 (7553):\penalty0 436--444, May 2015.
\newblock ISSN 1476-4687.
\newblock \doi{10.1038/nature14539}.
\newblock URL \url{https://doi.org/10.1038/nature14539}.

\bibitem[Li et~al.(2020)Li, Zhou, Xu, Luo, and Hu]{Li_2020}
YaoChong Li, Ri-Gui Zhou, RuQing Xu, Jia Luo, and WenWen Hu.
\newblock A quantum deep convolutional neural network for image recognition.
\newblock \emph{Quantum Science and Technology}, 5\penalty0 (4):\penalty0
  044003, jul 2020.
\newblock \doi{10.1088/2058-9565/ab9f93}.
\newblock URL \url{https://doi.org/10.1088/2058-9565/ab9f93}.

\bibitem[Mitarai et~al.(2018)Mitarai, Negoro, Kitagawa, and Fujii]{Grad0}
K.~Mitarai, M.~Negoro, M.~Kitagawa, and K.~Fujii.
\newblock Quantum circuit learning.
\newblock \emph{Phys. Rev. A}, 98:\penalty0 032309, Sep 2018.
\newblock \doi{10.1103/PhysRevA.98.032309}.
\newblock URL \url{https://link.aps.org/doi/10.1103/PhysRevA.98.032309}.

\bibitem[Monr\`as et~al.(2017)Monr\`as, Sent\'{\i}s, and Wittek]{QQ5}
Alex Monr\`as, Gael Sent\'{\i}s, and Peter Wittek.
\newblock Inductive supervised quantum learning.
\newblock \emph{Phys. Rev. Lett.}, 118:\penalty0 190503, May 2017.
\newblock \doi{10.1103/PhysRevLett.118.190503}.
\newblock URL \url{https://link.aps.org/doi/10.1103/PhysRevLett.118.190503}.

\bibitem[Nielsen \& Chuang(2011)Nielsen and Chuang]{NielsenChuang}
Michael~A. Nielsen and Isaac~L. Chuang.
\newblock \emph{Quantum Computation and Quantum Information: 10th Anniversary
  Edition}.
\newblock Cambridge University Press, USA, 10th edition, 2011.
\newblock ISBN 1107002176.

\bibitem[Oh et~al.(2021)Oh, Choi, Kim, and Kim]{9333906}
Seunghyeok Oh, Jaeho Choi, Jong-Kook Kim, and Joongheon Kim.
\newblock Quantum convolutional neural network for resource-efficient image
  classification: A quantum random access memory (qram) approach.
\newblock In \emph{2021 International Conference on Information Networking
  (ICOIN)}, pp.\  50--52, 2021.
\newblock \doi{10.1109/ICOIN50884.2021.9333906}.

\bibitem[Paparo et~al.(2014)Paparo, Dunjko, Makmal, Martin-Delgado, and
  Briegel]{CQ2}
Giuseppe~Davide Paparo, Vedran Dunjko, Adi Makmal, Miguel~Angel Martin-Delgado,
  and Hans~J. Briegel.
\newblock Quantum speedup for active learning agents.
\newblock \emph{Phys. Rev. X}, 4:\penalty0 031002, Jul 2014.
\newblock \doi{10.1103/PhysRevX.4.031002}.
\newblock URL \url{https://link.aps.org/doi/10.1103/PhysRevX.4.031002}.

\bibitem[P.Q.~Le \& Hirota(2011)P.Q.~Le and Hirota]{FRQI}
F.~Dong P.Q.~Le and K.~Hirota.
\newblock A flexible representation of quantum images for polynomial
  preparation, image compression, and processing operations.
\newblock \emph{Quantum Inf Process}, 10:\penalty0 63--84, 2011.
\newblock \doi{10.1007/s11128-010-0177-y}.

\bibitem[Preskill(2012)]{Sycamore3}
John Preskill.
\newblock Quantum computing and the entanglement frontier, 2012.

\bibitem[Qiskit(2017)]{qiskit}
Qiskit.
\newblock 2017.
\newblock URL \url{https://qiskit.org/textbook/preface.html}.

\bibitem[Rasmussen et~al.(2020)Rasmussen, Groenland, Gerritsma, Schoutens, and
  Zinner]{NToffoli1}
S.~E. Rasmussen, K.~Groenland, R.~Gerritsma, K.~Schoutens, and N.~T. Zinner.
\newblock Single-step implementation of high-fidelity n-bit toffoli gates.
\newblock \emph{Physical Review A}, 101\penalty0 (2), Feb 2020.
\newblock ISSN 2469-9934.
\newblock \doi{10.1103/physreva.101.022308}.
\newblock URL \url{http://dx.doi.org/10.1103/PhysRevA.101.022308}.

\bibitem[Sasaki \& Carlini(2002)Sasaki and Carlini]{QQ1}
Masahide Sasaki and Alberto Carlini.
\newblock Quantum learning and universal quantum matching machine.
\newblock \emph{Phys. Rev. A}, 66:\penalty0 022303, Aug 2002.
\newblock \doi{10.1103/PhysRevA.66.022303}.
\newblock URL \url{https://link.aps.org/doi/10.1103/PhysRevA.66.022303}.

\bibitem[Schuld et~al.(2014)Schuld, Sinayskiy, and Petruccione]{CQ3}
Maria Schuld, Ilya Sinayskiy, and Francesco Petruccione.
\newblock The quest for a quantum neural network.
\newblock \emph{Quantum Information Processing}, 13\penalty0 (11):\penalty0
  2567--2586, Nov 2014.
\newblock ISSN 1573-1332.
\newblock \doi{10.1007/s11128-014-0809-8}.
\newblock URL \url{https://doi.org/10.1007/s11128-014-0809-8}.

\bibitem[Schuld et~al.(2019)Schuld, Bergholm, Gogolin, Izaac, and
  Killoran]{Grad1}
Maria Schuld, Ville Bergholm, Christian Gogolin, Josh Izaac, and Nathan
  Killoran.
\newblock Evaluating analytic gradients on quantum hardware.
\newblock \emph{Physical Review A}, 99\penalty0 (3), Mar 2019.
\newblock ISSN 2469-9934.
\newblock \doi{10.1103/physreva.99.032331}.
\newblock URL \url{http://dx.doi.org/10.1103/PhysRevA.99.032331}.

\bibitem[Schuld et~al.(2020)Schuld, Bocharov, Svore, and Wiebe]{Schuld}
Maria Schuld, Alex Bocharov, Krysta~M. Svore, and Nathan Wiebe.
\newblock Circuit-centric quantum classifiers.
\newblock \emph{Physical Review A}, 101\penalty0 (3), Mar 2020.
\newblock ISSN 2469-9934.
\newblock \doi{10.1103/physreva.101.032308}.
\newblock URL \url{http://dx.doi.org/10.1103/PhysRevA.101.032308}.

\bibitem[Sent{\'i}s et~al.(2012)Sent{\'i}s, Calsamiglia, Mu{\~{n}}oz-Tapia, and
  Bagan]{QQ3}
G.~Sent{\'i}s, J.~Calsamiglia, R.~Mu{\~{n}}oz-Tapia, and E.~Bagan.
\newblock Quantum learning without quantum memory.
\newblock \emph{Scientific Reports}, 2\penalty0 (1):\penalty0 708, Oct 2012.
\newblock ISSN 2045-2322.
\newblock \doi{10.1038/srep00708}.
\newblock URL \url{https://doi.org/10.1038/srep00708}.

\bibitem[Sent\'{\i}s et~al.(2019)Sent\'{\i}s, Monr\`as, Mu\~noz Tapia,
  Calsamiglia, and Bagan]{QQ8}
Gael Sent\'{\i}s, Alex Monr\`as, Ramon Mu\~noz Tapia, John Calsamiglia, and
  Emilio Bagan.
\newblock Unsupervised classification of quantum data.
\newblock \emph{Phys. Rev. X}, 9:\penalty0 041029, Nov 2019.
\newblock \doi{10.1103/PhysRevX.9.041029}.
\newblock URL \url{https://link.aps.org/doi/10.1103/PhysRevX.9.041029}.

\bibitem[Shende \& Markov(2008)Shende and Markov]{NToffoli2}
Vivek~V. Shende and Igor~L. Markov.
\newblock On the cnot-cost of toffoli gates, 2008.

\bibitem[Shor(1997)]{Shor}
Peter~W. Shor.
\newblock Polynomial-time algorithms for prime factorization and discrete
  logarithms on a quantum computer.
\newblock \emph{SIAM Journal on Computing}, 26\penalty0 (5):\penalty0
  1484–1509, Oct 1997.
\newblock ISSN 1095-7111.
\newblock \doi{10.1137/s0097539795293172}.
\newblock URL \url{http://dx.doi.org/10.1137/S0097539795293172}.

\bibitem[Simon(1997)]{Simon}
Daniel~R. Simon.
\newblock On the power of quantum computation.
\newblock \emph{SIAM Journal on Computing}, 26\penalty0 (5):\penalty0
  1474--1483, 1997.
\newblock \doi{10.1137/S0097539796298637}.
\newblock URL \url{https://doi.org/10.1137/S0097539796298637}.

\bibitem[Smith et~al.(2016)Smith, Dherin, Barrett, and De]{implicitreg}
Samuel~L. Smith, Benoit Dherin, David G.~T. Barrett, and Soham De.
\newblock On the origin of implicit regularization in stochastic gradient
  descent.
\newblock In \emph{International Conference on Learning Representations},
  volume~29, 2016.
\newblock URL
  \url{https://proceedings.neurips.cc/paper/2016/file/d47268e9db2e9aa3827bba3afb7ff94a-Paper.pdf}.

\bibitem[Toffoli(1980)]{Toffoli}
T.~Toffoli.
\newblock Springer Berlin Heidelberg, Berlin, Heidelberg, 1980.

\bibitem[Wan et~al.(2017)Wan, Dahlsten, Kristjánsson, Gardner, and Kim]{QNN2}
Kwok~Ho Wan, Oscar Dahlsten, Hlér Kristjánsson, Robert Gardner, and M.~S.
  Kim.
\newblock Quantum generalisation of feedforward neural networks.
\newblock \emph{npj Quantum Information}, 3\penalty0 (1), Sep 2017.
\newblock ISSN 2056-6387.
\newblock \doi{10.1038/s41534-017-0032-4}.
\newblock URL \url{http://dx.doi.org/10.1038/s41534-017-0032-4}.

\end{thebibliography}

\appendix


\end{document}